\documentclass[useAMS]{mn2e}
\usepackage{epsfig}

\newif\ifAMStwofonts
\AMStwofontstrue

\def\be{\begin{equation}}
\def\ee{\end{equation}}

\def\gtsima{$\; \buildrel > \over \sim \;$}
\def\ltsima{$\; \buildrel < \over \sim \;$}
\def\prosima{$\; \buildrel \propto \over \sim \;$}
\def\gsim{\lower.5ex\hbox{\gtsima}}
\def\lsim{\lower.5ex\hbox{\ltsima}}
\def\simgt{\lower.5ex\hbox{\gtsima}}
\def\simlt{\lower.5ex\hbox{\ltsima}}
\def\simpr{\lower.5ex\hbox{\prosima}}

\def\Lya{Ly$\alpha$~}

\def\HI{\hbox{H$\scriptstyle\rm I\~$}}
\def\nHI{{\rm HI~}}

\def\nH{{\rm H}}

\def\nHeII{{\rm HeII~}}

\def\HI{\hbox{H~$\scriptstyle\rm I\ $}}

\def\sph{{\tt sph~}}

\def\mfb{{\tt mfb~}}

\def\ie{{\frenchspacing\it i.e. }}
\def\eg{{\frenchspacing\it e.g. }}
\def\etal{{\it et al.~}}

\title[The \Lya Forest Around High-$z$ Galaxies]{The \Lya 
Forest Around High Redshift Galaxies}

\author[M. Bruscoli, A. Ferrara, S. Marri, R. Schneider, A.
Maselli, E. Rollinde \& B. Aracil]{M. Bruscoli$^1$, 
A. Ferrara$^{2}$, S. Marri$^2$, R. 
Schneider$^3$, A. Maselli$^1$, 
\newauthor E. Rollinde$^4$ \& B. Aracil$^4$\\
$^1$Dipartimento di Astronomia, Universit\`a degli Studi 
di Firenze, L.go E. Fermi 2, 50125 Firenze, Italy\\
$^2$SISSA/International School for Advanced Studies, Via Beirut 2-4, 34014 Trieste, Italy\\
$^3$Osservatorio Astrofisico di Arcetri, L.go E. Fermi 5, 50125
Firenze, Italy\\
$^4$Institut d'Astrophysique de Paris, 98bis Boulevard d'Arago, 75014 Paris, France\\}
\pagerange{\pageref{firstpage}--\pageref{lastpage}}
\pubyear{2001}
\begin{document}

\maketitle
\label{firstpage}

\begin{abstract}
Motivated by the relative lack of neutral hydrogen around Lyman Break Galaxies
deduced from recent observations, we investigate the properties of the \Lya forest
around high redshift galaxies. The study is based on improved numerical SPH simulations
implementing, in addition to standard processes, a new scheme for multiphase
and outflow physics description. 
Although on large scales our simulations reproduce a number
of statistical properties of the IGM (because of the
small filling factor of shock-heated gas), they underpredict the 
\Lya optical depth decrease inside 1 Mpc~$h^{-1}$ of the galaxies 
by a factor of $\approx 2$. We interpret this result  as due
to the combined effect of infall occurring along the filaments, which prevents efficient halo 
gas clearing by the outflow, and the insufficient increase of
(collisional) hydrogen ionization produced by the temperature increase
inside the hot, outflow-carved bubble. Unless an observational selection bias 
is present, we speculate that local photoionization could be the only 
viable explanation to solve the puzzle.

\end{abstract}
 

\section{INTRODUCTION}
Galaxies form from the intergalactic medium (IGM), process such gas
into stars, and possibly re-eject a fraction of it, enriched by 
nucleosynthetic products, back into the intergalactic space via
powerful supernova-driven outflows (Mac Low \& Ferrara 1999;
Ferrara, Pettini \& Shchekinov 2000;
Madau, Ferrara \& Rees 2001; Scannapieco, Ferrara \& Madau 2002; 
Theuns \etal 2002). 
The energy deposition connected to these processes is expected
to leave at least some detectable imprints on the physical state 
of the IGM. Thus, it is conceivable that such signatures can be studied
through QSO absorption line experiments. Naively, the presence of
hot outflowing gas should result primarily in two effects:
[i] a decrease of the gas density and [ii] an increase of the 
temperature caused by shock-heating (acting in conjunction with 
photo-heating by the UV background) in a large (several hundreds
kpc) region around the perturbing galaxy. Both these occurrences would
imply an increasingly more transparent Ly$\alpha$ forest when 
approaching the galaxy, \ie a galactic proximity effect.
Quantitative confirmation 
of this scenario has faced tremendous difficulties, standing 
the complications of the physics of star formation, explosions and
metal mixing in multiphase media. Hence, most simulations to date 
had to rely on {\it ad hoc} recipes for such processes. \\
Nevertheless, these ideas have stimulated the first   
challenging observations aimed at detecting the imprints of 
galaxy-IGM interplay.
Adelberger \etal (2002, A02) obtained high resolution spectra of 8 bright 
QSOs at $3.1<z<4.1$ and spectroscopic redshifts for 431 Lyman-break 
galaxies (LBGs) at lower redshifts. By comparing the positions of the 
LBGs with the Ly$\alpha$ absorption lines in QSO spectra, indeed
they conclude that within $\approx 0.5 h^{-1}$ (comoving) Mpc of the galaxies 
little \HI              is present; on the contrary, between 1 
and 5$h^{-1}$ Mpc an \HI excess with respect to the IGM mean is detected. 
This simple interpretation might be at odd with the results of a  
VLT/UVES study of the \Lya forest in the vicinity of the LBG MS1512-cB58 
showing the opposite trend (Savaglio \etal 2002), \ie an 
absorption excess close to the galaxy. \\
Numerical simulations have also noticeable difficulties reproducing
A02 results as discussed by Croft \etal (2002) and Kollmeier \etal (2002);
however, these studies lack a self-consistent treatment of 
multiphase gas structure and/or outflow dynamics. 
Here we revisit A02 results through SPH simulations 
(Marri \etal  2003) that
implement a new scheme for multiphase hydrodynamics and, more importantly, 
a physically meaningful outflow treatment; the full description  
of the code and of the tests made are given in Marri \& White (2002).
We then derive synthetic absorption line spectra along lines of
sight randomly traced through the simulation box at $z\approx 3$
and compare them directly with A02 data to investigate  
galactic feedback effects on the IGM.

\section{Simulations and data analysis}
\label{sect2}
We have performed hydrodynamic simulations for a 
$\Lambda$CDM cosmological model with $\Omega_{0}=0.3$, $\Omega_{\Lambda}=0.7$, 
$\Omega_{b}=0.04$ and $h=0.7$ km s$^{-1}$ Mpc$^{-1}$. The initial power
spectrum is cluster-normalized ($\sigma_{8}=0.9$); periodic boundary conditions are
adopted. 
We first obtained a set of low-resolution runs ($64^{3}$ particles for both gas
and dark matter) in a $7h^{-1}$ comoving Mpc cube;
these runs serve as a guide for more computationally expensive runs and 
for testing purposes. We consider a first model where the IGM 
multiphase structure and galaxy outflows are deliberately ignored
and a second one which includes a description of both these physical effects. 
We will refer to these runs as \sph and \mfb, respectively,  with the
same meaning (and parameters) adopted in the description of the 
low-resolution       
$\Lambda$CDM test problem described in Marri \& White (2002). 
The high-resolution run, on which the main results of the present
analysis are based,
is a $128^{3}$ particles simulation in a $10.5h^{-1}$ comoving Mpc
box. For this run we only studied the full \mfb model.\\ 
Softening lenghts are fixed both in physical and comoving coordinates 
as required in GADGET (Springel, Yoshida \& White 2001). Gas (dark matter) 
gravitational softening is approximately $6h^{-1}$~kpc ($8h^{-1}$~kpc) comoving 
and $3h^{-1}$~kpc ($4h^{-1}$~kpc) physical in
the high-res        case and scales according to particle number and
box size in the low-res run.
In all runs we include the effects of a UV background produced by QSOs and 
filtered through
the IGM,  whose shape and amplitude are taken from Haardt \& Madau (1996). 

\subsection{Synthetic Ly$\alpha$ forest spectra}
To allow a direct comparison of simulation results with observational data
we construct synthetic \Lya spectra from simulation outputs 
at redshift $z=3.17$ 
($z=3.27$) for the low-res (high-res) runs.
In the simulation box we randomly trace 60 lines-of-sight (LOS) parallel 
to the $x$-axis, each of which is discretized
into $N_{pix}=1024$ pixels. In order to assign to each 
pixel a value for the hydrogen density, temperature and peculiar velocity
we perform a standard SPH smoothing using the 32 closest neighbour
SPH particles to the LOS pixel position. 
The neutral hydrogen density
is derived using the code CLOUDY94\footnote{http://nimbus.pa.uky.edu/cloudy/} 
on each pixel, adopting the same UVB as in the simulation. 
The transmitted flux due to Ly$\alpha$ absorption in the IGM is 
$\propto e^{-\tau}$, where $\tau$ is the optical depth along the considered LOS.
The Hubble velocity $v_{H}$  varies in the range (0,$v_{H}^{max}$), where 
the maximum value is set by the box size.
The contribution to $\tau$ at the observed frequency corresponding to $v_H(k)$,
where $1\le k \le 1024$ is the pixel index, is given by 
\begin{equation}
\tau[v_{H}(k)]=\frac{\Delta x \sigma_{0}f\lambda_{0}}{\pi^{1/2}} \sum_{i=1}^{N_{pix}}\frac{ n_{\nHI}(i)}{
{b(i)}}\Phi_{g}[v_{H}(k)-v(i)] ,
\label{eq1}
\end{equation}
where $\Delta x$ is the pixel size, $\sigma_{0}$ is the Ly$\alpha$ cross section, $f$ is the
Ly$\alpha$ oscillator strength, 
$\lambda_{0}$ is the Ly$\alpha$ wavelength, $n_{\HI}(i), b(i)$ and $v(i)=v_{H}(i)+v_{pec}(i)$
are the \HI number density, the Doppler parameter, and the total velocity in the
pixel $i$, respectively; $\Phi_{g}$ is a gaussian line profile and $v_{pec}$ is the peculiar
velocity. For additional discussion on this formula, see \eg Rauch, Haehnelt \& Steinmetz (1997a).\\
To compare observations and simulations at best it is necessary to degrade the synthetic spectra  
to account for the uncertainties affecting the observed spectra. 
We perform such procedure through the following steps: {\it i}) continuum normalization 
at the highest flux value in each spectrum; 
{\it ii}) convolution with the instrumental profile; {\it iii}) sampling due to spectrograph 
spatial resolution; {\it iv}) addition of instrumental noise (see Rauch 
\etal  1997b;
Theuns \etal  1998; McDonald \etal  2000; Petry \etal  2002).
As most of the data we compare with are taken with the Keck/HIRES
spectrograph we adopt the following instrumental characteristics:
FWHM $= 8.0$ km s$^{-1}$, pixel spectral resolution $\Delta \lambda=0.05$\AA~ 
and signal-to-noise ratio S/N=50. The low (high) resolution box 
has a wavelength extent of $\Delta \lambda_{box}=13.46$\AA~
($\Delta \lambda_{box}=20.83$\AA); hence, degraded spectra in 
the two cases are made of 269 and 417 pixels, respectively.

\section{Results}
\label{sect3}
\begin{figure}
\centerline{{
\epsfig{figure=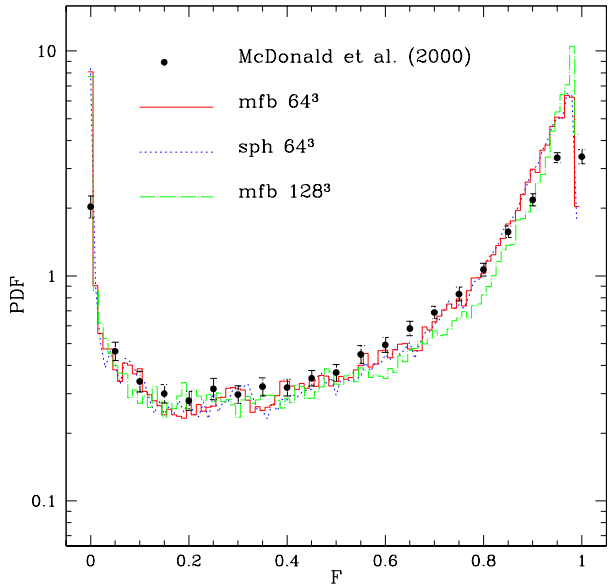, height=9cm}
}}
\caption{{\protect\footnotesize {Probability distribution function of transmitted 
flux at $z=3.17$ for the low-res        simulations 
\mfb (solid line) and \sph (dotted) runs, and at $z=3.27$ for the high
resolution \mfb run (dotted-dashed). The points represent the observational data
(McDonald \etal 2000) at $z=3$. }}} 
\label{fig1}
\end{figure}
As a general sanity check of the simulations we have first confronted
the  simulated statistical properties of the Ly$\alpha$ forest with 
the observed ones. The most obvious comparison involves the
probability distribution function (PDF) of the transmitted flux. 
In Fig.\ref{fig1} we plot the PDF as a function
of the flux $\langle F \rangle = e^{-\tau}$ for the low-res        
(both \mfb and \sph runs) and high-res       
simulations (\mfb case only) and compare them with the
observational data of McDonald \etal  (2000).
The general agreement is quite good through all the flux range 
and for all three cases.
The discrepancy between simulations and data at high fluxes  
is  probably due to the uncertainties in the data continuum fitting
(McDonald \etal  2000; Croft \etal  2002).
Surprisingly, it appears that the inclusion of multiphase
and outflow physics, not considered in the pure \sph run, does not
affect the distribution in a sensible manner. In other words,
galactic outflows leave the \Lya forest unperturbed.
This result is in agreement with that recently found by Theuns \etal  (2002),
who interpreted it as an indication that galactic outflows tend to propagate 
preferentially in the voids leaving the \Lya absorbing filaments virtually 
unaffected. 
As an additional check we have calculated the Doppler parameter, $b$, 
and the \HI column density distribution, 
and compared them with two QSOs observations at $z_{em}=3.38$ and
$z_{em}=3.17$ by 
Hu \etal  (1995).   
The values of $b$ and $N_{\nHI}$ for each absorber in the synthetic 
spectra have been derived using the fitting programm AUTOVP.
The experimental $b$ distribution is well reproduced both by the 
\sph and \mfb (low-res + high-res) runs at similar quality level
(simulated and observed distributions peak both around $b\sim 20$~km~s$^{-1}$). 
However, simulated spectra
tend to slightly overproduce lines with $10< b <22$~km~s$^{-1}$ and under-predict  
lines with $25 < b < 50$~km~s$^{-1}$. A similar effect has already been
noted by Theuns \etal (1998). These authors propose a number of possible 
explanations for this behavior: physical (\nHeII reionization, radiative transfer effects),
numerical (resolution) and related to data analysis (fitting procedure). Our
results seem to indicate that numerical artifacts should not be the dominant factor.
As for the \HI column density distribution, the \mfb case seems to  
reproduce the observational data for $\log N_{\nHI}>15.5$ somewhat
better; for smaller values of $N_{\nHI}$ the differences between the two runs 
are negligible.\\
\begin{figure}
\centerline{{
\epsfig{figure=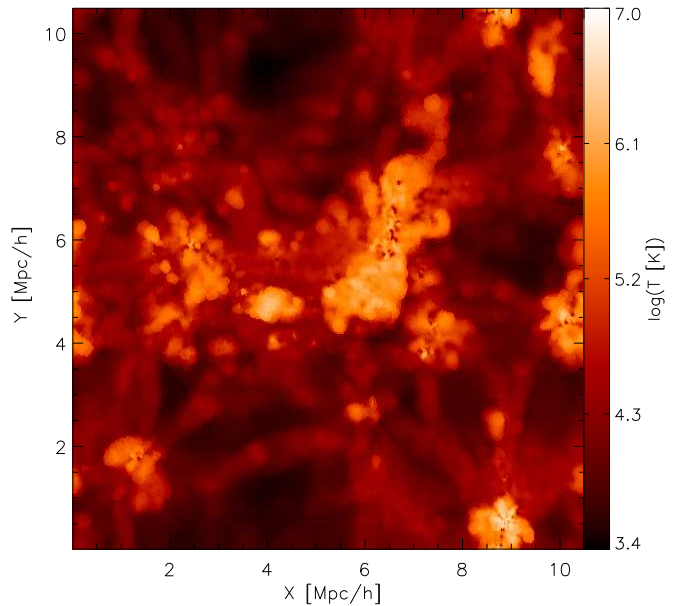, height=10cm}}}
\caption{{\protect\footnotesize {Temperature map in a slice 
through the high-res simulation box ($z = 3.27$). The color
bar shows the values of $\log T$.
}}} 
\label{fig2}
\end{figure}
In Fig.\ref{fig2} we show a temperature map from a slice through
the high-res box  at $z=3.27$. Hot bubbles ($T\simeq 10^6$~K) of 
shocked gas produced by outflows around the parent galaxies are clearly 
apparent. 
Their sizes range from $\approx 0.5$ Mpc $h^{-1}$ to $\approx 2$ Mpc $h^{-1}$, and their
shape appears in some case rather jagged as a result of the interaction
with the inhomogeneous ambient medium. The volume filling factor of gas with 
temperature above $10^5$~K is 14\%. 
The internal structure of the bubbles can be inspected more quantitatively in 
Fig.\ref{fig3}, where the comparison between \sph and \mfb is shown 
(low-res case) for various physical quantities (hydrogen density, $n_{\nH}$, 
ionization fraction, $x_{\nHI}$,
gas temperature, T, and peculiar velocity, $v_{pec}$) along the LOS through
the center of the most massive galaxy $M = 1.5\times 10^{11} M_\odot {\rm h}^{-1}$
in these simulations; the galaxy position corresponds to the
density peak at $x\approx 6.8$~Mpc~$h^{-1}$. 
In both simulations the star formation rate in this galaxy is 5.6 $M_\odot {\rm
yr}^{-1}$ (18 $M_\odot {\rm yr}^{-1}$) for the \mfb (\sph) models.  
A striking result of the comparison between the two models
shows that, although outflows are able to heat the halo/IGM
gas up to high temperatures out to more than 1 Mpc~$h^{-1}$ from the galaxy,
they do not seem able to modify its density structure in a sensible
way.
Hence the density in the surroundings remains high and close to
that set up by the process of galaxy formation. Close to the galactic center 
the outflow peak speed is about $130$~km~s$^{-1}$, but this value rapidly 
decreases as kinetic energy is used to counteract the pressure of
intergalactic accreting gas, raining onto the galaxy at essentially 
the escape speed of the system, roughly $150$~km~s$^{-1}$. 
The stalling radius is seen at the zero-crossing
of $v_{pec}$, approximately 0.3~Mpc~$h^{-1}$ away from the outflow source. 
The relative insensitivity of the density to the SN energy injection 
can be interpreted as the fact that the outflow velocities are lower 
than the escape speeds and hence the flow is confined by inflow. 
\begin{figure}
\centerline{{
\epsfig{figure=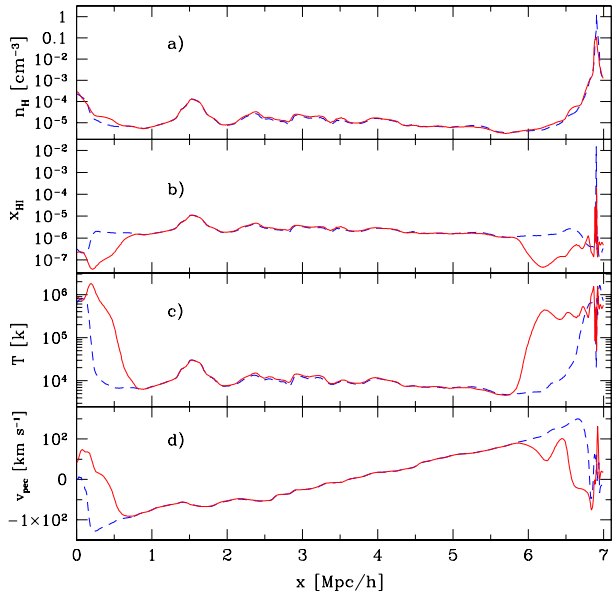, height=9cm}}}
\caption{{\protect\footnotesize {Comparison between  
\sph (dashed line) and \mfb (solid line) low-res runs along 
the LOS through the center of the most massive galaxy
($M=1.5\times 10^{11}M_{\odot}h^{-1}$) whose position corresponds
to the
density peak at $x\approx 6.8$~Mpc~$h^{-1}$: {\it a)}
hydrogen density, {\it b)} neutral hydrogen fraction, {\it c)} gas temperature, {\it d)} 
gas peculiar velocity.
 }}} 
\label{fig3}
\end{figure}
Also, we note that the outflow velocities we find are lower than those inferred 
by A02, $\approx 600$ km s$^{-1}$ from LBGs.\\
Fig.\ref{fig4} shows the mean Ly$\alpha$ forest flux averaged on pixels on
different LOS at different distances, $\Delta r$, from the galaxy centers in the simulation box,
following the same experimental procedure as in A02: 
\begin{equation}
\big{<}F(\Delta r)\big{>}=\frac{1}{N(\Delta r)} \sum_{i=1}^{N(\Delta r)}F_{i},
\end{equation}
where $N(\Delta r)$ is the number of pixels that fall in the bin $\Delta r$, 
and $F_{i}$ is the transmitted flux in the pixel $i$.
The synthetic data, normalized to the observational ones, are plotted 
only out to a distance $\Delta r \approx 3.81$~Mpc~$h^{-1}$ for the low-res
simulations ($\Delta r \approx 5.32$~Mpc~$h^{-1}$ for high-res one) due to the 
periodic boundary conditions of the simulation. 
We first compare the two low-res \sph and \mfb cases. 
The difference between these curves is negligible and on average 
the Ly$\alpha$ absorption is similar: this is expected after the
results analyzed in Fig. \ref{fig1}. 
In addition both simulations are in agreement with the high-res one, thus
ensuring that numerical convergence has been reached and results are not
affected by spurious effects. However, these models fail 
in reproducing the observed trend inside  1 Mpc~$h^{-1}$. The simulated
\Lya forest seems to be much more opaque than the observed one, with a 
maximum $\tau$ discrepancy of about a factor 2.2. In order to assess if 
the disagreement could be reduced by a more efficient outflow clearing
of the halos of smaller systems, we excluded the contribution of most 
massive galaxies ($M>1.5\times10^{10}h^{-1}M_{\odot}$) to the flux.
However, as it is seen from the Figure, this does not solve the problem.
As a final resort, we have randomized the position of galaxy centers in
the box to account for the reported experimental error determination.
To this aim we have added a gaussian random displacement to the redshift of each
galaxy with r.m.s. $\sigma_z =0.002$, which represents the 1-$\sigma$
error of the measure. This procedure improves the result, as now galaxy centers
do not perfectly coincide with density peaks, but to an extent insufficient
to explain the data. 
\begin{figure}
\centerline{{
\epsfig{figure=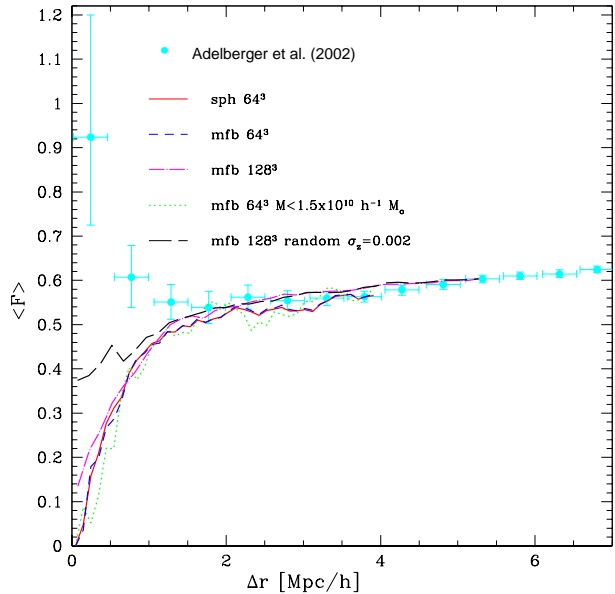, height=9cm}}}
\caption{{\protect\footnotesize {Mean \Lya forest flux
at different distances $\Delta r$ from the galaxies in the
simulation box. Synthetic results are shown for five cases:
low-res \sph (solid line) and \mfb (dashed); low-res \mfb with least massive
($M<1.5\times10^{10}h^{-1}M_{\odot}$) galaxies only (dotted); 
high-res \mfb (dot-dashed); high-res \mfb with randomized galaxy centers (long-dashed).
The points are the observational results by A02 for LBGs. 
Normalization to the data at large $\Delta r$ has been performed (see
text).
 }}} 
\label{fig4}
\end{figure}

%\begin{table}
%\centerline{Table 1: Reionization parameters}
%\begin{center}
%\begin{tabular}{|cccccccc|}
%\hline \hline
%Run&$\Delta z_{s}$$^{(a)}$&$z_{i,s}$$^{(b)}$&$\tau_{s}$$^{(c)}$&
%$z_{i,p}$$^{(d)}$ &$x_{e,s}(z_{i,p})$$^{(e)}$&$\frac{\Delta C_\ell}{C_\ell}$&$\frac{\Delta P_\ell}{P_\ell}$\\ \hline \hline
%$A$& 15&$ 10.9$&0.080&17.1&0.46&$\lsim 0.01$&$\lsim 0.06$ \\ \hline
%
%$B$& 17&$ 8.3$&0.059&13.9&0.51&$\lsim 0.007$&$\lsim 0.1$ \\ \hline
%\end{tabular}
%\label{tab1}
%\caption{{\protect\footnotesize{$^{(a)}$Duration of reionization; 
%$^{(b)}$Redshift of 
%complete reionization (simulated); $^{(c)}$Thomson Optical Depth (simulated);
%$^{(d)}$Redshift of complete reionization (prompt); $^{(e)}$Hydrogen ionization fraction 
%at $z_{i,p}$ (simulated).}}} 
%\end{center}
%\end{table}

\section{Conclusions}
\label{sect4}
Motivated by the recent observational results of A02, 
we have studied, with the help of a set of cosmological simulations including 
star formation in multiphase gas and outflows from galaxies, the effects
of galaxy formation/activity on the properties of the surrounding \Lya forest.
Although on  large scales our simulations can reproduce remarkably well a number
of statistical properties of the IGM, they fail to predict the observed
\Lya flux increase in regions close to the galaxies themselves. The success
can be ascertain to two concomitant effects: (i) outflows preferentially 
expand in regions of low-density (voids) thus preserving the filaments
responsible for the \Lya absorbing network; (ii) the hot bubbles 
fill a relatively small fraction of the cosmic volume ($\approx 14$\%
in our simulations). 
Support to the first hypothesis emerges also from an inspection of   
the velocity field in the surrounding of the most 
massive galaxy in the simulation:  
the inflow of gas from low density
regions is blocked by the bubble expansion, but it proceeds basically
unimpeded along the filaments (see Marri \etal 2003).\\
Much more puzzling is instead
the reason for the opacity excess (with respect to real data) we see in 
the simulation in the inner Mpc~$h^{-1}$. Apparently, simulated outflows do not  
carry sufficient momentum to disperse the density peak created in the vicinity 
of the galaxy as a leftover of its formation. Also, a large fraction of the 
outflow energy is used 
to counteract the infalling gas ram pressure which tends to pile up the 
gas into the galaxy. The temperature increase close to galaxies amplifies
the magnitude of the collisional ionization rate, which becomes larger than
the equivalent photoionization rate for $T\simgt 10^5$~K. In this case 
the \HI neutral fraction $x_{\nHI}$ is independent of gas density and remains at roughly
the same level as in the general IGM. Therefore, collisional ionization is not
sufficient to balance the opacity increase induced by the density raise and the
transmitted flux drops accordingly to the latter.
What are the possible alternative explanations for the observed flux trend?
The presence of a bias in the data produced by a preferential selection effect 
of low \Lya absorption LBGs has already been suggested by Croft \etal (2002).
Another possibility is provided by photoionization, particularly if one recalls  
the recent results  by Steidel, Pettini \& Adelberger (2001),
who detected flux beyond the Lyman limit (with significant residual flux
at $\lambda <  912$~\AA) in a composite spectrum of 29 LBGs at $z = 3.4$, a
clue of a conspicuous escape probability of ionizing radiation from these objects.
As mentioned above, the A02 data would require an optical depth (or equivalently
a $x_{\nHI}$, assuming a prescribed density profile) a factor $\approx 2-3$ smaller.
If this can be achieved with the ionizing flux coming from galaxies must be proved with 
detailed radiative transfer calculations which are currently ongoing (Maselli \etal 2003)
The first attempt using simplified analytical and/or post-processing techniques 
to account for this effect (Croft \etal 2002, Kollmeier \etal 2002) have yielded so
far negative answers, but a fully self-consistent, physically accurate description
of the problem is awaited in order to draw a final conclusion.

%END%END%END%END%END%END%END%END%END%END%END%END%END%END%END%END%END%END
\bigskip
This work was partially supported by the Research and Training Network
`The Physics of the Intergalactic Medium' set up by the European Community
under the contract HPRN-CT2000-00126 RG29185.
MB thanks P.Petitjean for discussions and hospitality
at IAP. We are grateful to S. Bianchi for help with AUTOVP and
discussions.

\label{lastpage}
\newpage
\end{document}